\documentclass[12pt,preprint]{aastex}
\usepackage{epsfig}
\usepackage{natbib}
\usepackage{graphicx}
\usepackage{slashbox}
\usepackage{multirow}
\usepackage{lscape}
\usepackage{mathrsfs,amssymb}
\usepackage{amsmath}
\newcommand       \Angstrom     {\,{\rm \AA}}

\newcommand       \cm           {\,{\rm cm}}

\newcommand       \gtsim        {\gtrsim}

\newcommand       \mum          {\,{\rm \mu m}}

\newcommand       \simali       {\sim\,}
\newcommand       \magni    {\,{\rm mag}}

\newcommand  \fbump    {f_{\rm b}} 
\newcommand  \fdib      {f_{\rm d}}
\newcommand  \WDIBp  {W^{\prime}_{\rm DIB}}
\newcommand  \Wbumpp  {W^{\prime}_{\rm 2175}}
\shorttitle{DIBs vs. the 2175$\Angstrom$ Extinction Bump}
\title{
 \vspace*{1.0em}
A Tale of Two Mysteries in Interstellar Astrophysics: 
The 2175$\Angstrom$ Extinction Bump and Diffuse Interstellar Bands
}

\author{F.Y. Xiang\altaffilmark{1,2}, Aigen Li\altaffilmark{2} 
                                and J.X. Zhong\altaffilmark{1,2}}
\altaffiltext{1} {Department of Physics, Xiangtan University, 
                  411105 Xiangtan, Hunan Province, China;
                  {\sf fyxiang, jxzhong@xtu.edu.cn}}
\altaffiltext{2} {Department of Physics and Astronomy,
                  University of Missouri,
                  Columbia, MO 65211, USA;
                  {\sf lia@missouri.edu}}

\begin{document}

\begin{abstract}
The diffuse interstellar bands (DIBs) are ubiquitous absorption 
spectral features arising from the tenuous material in the space 
between stars -- the interstellar medium (ISM). Since their first 
detection nearly nine decades ago, over 400 DIBs have been observed 
in the visible and near-infrared wavelength range in both the Milky Way 
and external galaxies, both nearby and distant. However, the identity 
of the species responsible for these bands remains as one of the most 
enigmatic mysteries in astrophysics. 
An equally mysterious interstellar spectral signature 
is the 2175$\Angstrom$ extinction bump, 
the strongest absorption feature observed in the ISM. 
Its carrier also remains unclear 
since its first detection 46 years ago. 
%
%Recently, Cox et al. reported that in the Small Magellanic Cloud 
%the presence or absence of DIBs appears to be related to the presence 
%or absence of the 2175$\Angstrom$ extinction bump. 
%They argued that their carriers may be subject to the same formation 
%and destruction processes in the ISM. 
%
Polycyclic aromatic hydrocarbon (PAH) molecules
have long been proposed as a candidate for DIBs 
as their electronic transitions occur in the wavelength
range where DIBs are often found.
In recent years, the 2175$\Angstrom$ extinction bump
is also often attributed to the $\pi$--$\pi^{*}$ transition in PAHs. 
If PAHs are indeed responsible for both the 2175$\Angstrom$ 
extinction feature and DIBs, their strengths may correlate. 
We perform an extensive literature search for 
lines of sight for which both the 2175$\Angstrom$ 
extinction feature and DIBs have been measured. 
Unfortunately, we found no correlation between 
the strength of the 2175$\Angstrom$ feature and 
the equivalent widths of the strongest DIBs. 
A possible explanation might be that DIBs are produced 
by small free gas-phase PAH molecules and ions, 
while the 2175$\Angstrom$ bump is mainly from large PAHs
or PAH clusters in condensed phase so that there is 
no tight correlation between DIBs 
and the 2175$\Angstrom$ bump. 
%
%
%
%Here we report a close correlation between 
%the strength of the 2175NE extinction bump 
%and the strengths of eight strong DIBs 
%in the visible wavelength range for a large number 
%of Galactic ISM regions. 
%This indicates that the enigmatic carriers 
%of both the 2175NE extinction bump and DIBs 
%are probably related and share a common carrier.   
%
\end{abstract}
\keywords {dust, extinction; ISM: lines and bands; ISM: molecules}

\section{Introduction\label{sec:intro}}
The diffuse interstellar bands (DIBs) show up as absorption troughs 
in the optical/near infrared (IR) wavelength range
against the continuous spectra of hot background stars 
which lie behind diffuse interstellar clouds. 
The strength of a DIB is often measured in terms of 
the ``equivalent width'' (which reflects the wavelength-integrated 
area of the DIB; see Fig.\,1). 
As already recognized 76 years ago (Merrill et al.\ 1934), 
the DIB strengths generally increase with the amount of 
interstellar material between the background star and the observer, 
revealing their nature of interstellar origin.

%The year of 2009 marks the 75th anniversary of the first
%recognition of the interstellar nature of 
%the diffuse interstellar bands (DIBs; Merrill et al.\ 1934).
Since their first detection in 1919 by Heger (1922),
to date over four hundred DIBs have been observed 
in both Galactic and extragalactic sources 
(Snow 2002; Sarre 2006; Hobbs et al.\ 2008, 2009), 
including the Large and Small Magellanic Clouds 
(Ehrenfreund et al.\ 2002; Cox et al.\ 2006, 2007a;
Welty et al.\ 2006), 
M31 (Cordiner et al.\ 2008b, 2011),
M33 (Cordiner et al.\ 2008a),
starburst galaxies (Heckman \& Lehnert 2000),
intervening absorption systems toward quasars
(Ellison et al.\ 2008) and damped Ly$\alpha$ absorbers 
at cosmological distances (DLAs; Junkkarinen et al.\ 2004, 
York et al.\ 2006a, Lawton et al.\ 2008).
It is interesting to note that both the 2175$\Angstrom$
extinction bump (see below) and three weak DIBs 
at $\lambda$4430$\Angstrom$, $\lambda$5705$\Angstrom$, 
and $\lambda$5780$\Angstrom$ have been detected 
in the quasar intervening DLA system at $z_{\rm abs}=0.524$
toward AO\,0235+164 
(Junkkarinen et al.\ 2004, York et al.\ 2006a, Lawton et al.\ 2008).
   
The 2175$\Angstrom$ extinction bump shows up as 
the most prominent spectral feature in the interstellar 
extinction curve -- the wavelength dependence of the interstellar 
dust obscuration (see Fig.\,1). 
It is ubiquitously seen in the Milky Way (see Draine 1989)
and has recently also been detected in distant galaxies 
at redshifts $z>2$ (El{\'{\i}}asd{\'o}ttir et al.\ 2009;
Noll et al.\ 2007, 2009; Prochaska et al.\ 2009; 
Liang \& Li 2009, 2010).
Its detection at $z<2$ has been more frequently reported 
(e.g. see Motta et al.\ 2002, Wang et al.\ 2004, Noll \& Pierini 2005, 
York et al.\ 2006b, Inoue et al.\ 2006, Srianand et al.\ 2008, 
Conroy et al.\ 2010, Zhou et al.\ 2010, Jiang et al.\ 2011).
The most striking characteristics of the 2175$\Angstrom$ 
extinction bump are the invariant central wavelength and 
variable bandwidth: its peak position at 2175$\Angstrom$ 
is remarkably constant while the bandwidth varies from one 
line of sight to another (see Whittet 2003).

The exact nature of the carriers of DIBs 
remains as one of the most enigmatic mysteries 
in astrophysics (see Sarre 2006),
although polycyclic aromatic hydrocarbon (PAH) 
molecules seem to be a promising candidate 
(see Crawford et al.\ 1985, L\'eger \& d'Hendecourt 1985, 
van der Zwet \& Allamandola 1985, Salama et al.\ 1996, 1999, 2011). 
Other candidate materials such as fullerenes
(Ehrenfreund \& Foing 1996) have also received much attention.

The 2175$\Angstrom$ extinction bump also remains unidentified
since its first detection over four decades ago by Stecher (1965). 
It was generally attributed to the $\pi$--$\pi^{*}$ transition in 
$sp^2$ aromatic carbon dust such as graphite (Stecher \& Donn 1965)
or PAHs (Joblin et al.\ 1992, Li \& Draine 2001, Duley 2006).
%The presence of PAHs in interstellar space is revealed by
%the ubiquitous infrared (IR) emission features at 3.3, 6.2, 7.7,
%8.6, 11.3 and 12.7$\mum$ (L\'eger \& Puget 1984, 
%Allamandola et al.\ 1985).

Are the carriers of DIBs and the 2175$\Angstrom$ extinction bump related? 
more specifically, are PAHs -- a family of various aromatic species
which are thought to be widespread in interstellar space 
as revealed by the ubiquitous ``Unidentified IR (UIR)'' 
emission features at 3.3, 6.2, 7.7, 8.6, 11.3 and 12.7$\mum$ 
(L\'eger \& Puget 1984, Allamandola et al.\ 1985, 
Smith et al.\ 2007, Tielens 2008) --
responsible for both DIBs and the 2175$\Angstrom$ extinction bump? 

One way to test this hypothesis is to examine whether the strengths
of DIBs and the 2175$\Angstrom$ bump are correlated. %\footnote{%
%  Cox et al.\ (2007a) found that in the Small Magellanic Cloud,
%  the presence or absence of DIBs appears to be related to
%  the presence or absence of the 2175$\Angstrom$ extinction bump.
%  In the Galaxy, some lines of sight also exhibit both weak DIBs
%  and a weak extinction bump
%  (e.g. HD\,23532, Witt et al.\ 1981; HD\,29647, Adamson et al.\ 1991; 
%        HD\,62542, Snow et al.\ 2002, \'Ad\'amkovics et al.\ 2005)
%  on a per unit reddening basis, when compared with typical
%  Galactic sightlines. 
%  }
%%
We have initiated a program to explore the possible 
correlations between the DIBs and various interstellar extinction 
properties. In this work we perform a correlation study
of the DIB equivalent widths (EWs) and the 2175$\Angstrom$ 
extinction bump strengths, both are normalized by 
the interstellar reddening $E(B-V)$,\footnote{%
  $E(B-V)\equiv A_B-A_V = A_V/R_V$ is the difference between
  the $B$ band extinction ($A_B$) and
  the $V$ band extinction ($A_V$),
  where $R_V\equiv A_V/E(B-V)$ is the total-to-selective extinction ratio.
  }
for 84 sightlines 
of which the DIB equivalent widths, 
the 2175$\Angstrom$ bump strengths, 
%(or the total-to-selective extinction ratios $R_V\equiv A_V/E(B-V)$),
and $E(B-V)$ are available in the literature.
We use the equivalent width (EW) of a DIB to characterize 
its absorption strength, which measures the area of the DIB
absorption profile above the continuum 
(see the shaded area in the inserted panel in Fig.\,1).
In \S2 and \S3 we will discuss what would be the most 
reasonable way to characterize the strength of 
the 2175$\Angstrom$ extinction bump.

\section{DIBs vs. the 2175$\Angstrom$ Bump: 
         Where Do We Stand?\label{sec:history}}
Whether the carriers of DIBs are related to
the 2175$\Angstrom$ extinction feature has been
a topic of extensive research for over three decades.
Contradicting conclusions have been drawn in the literature.
As summarized in the following paragraphs,
this seems to be largely caused by the inconsistencies 
in determining the 2175$\Angstrom$ bump strength. 
 
Nandy \& Thompson (1975) first speculated that
      the 2175$\Angstrom$ extinction bump may be
      correlated with the strength of 
      the $\lambda$4430$\Angstrom$ DIB.
      They derived the ``equivalent widths'' of
      the bump (to represent the bump strengths)
      of $>$\,60 sightlines from the interstellar reddening
      $E(2190\Angstrom-2500\Angstrom)$.\footnote{%
        By assuming a bump width of 360$\Angstrom$ 
        and a mean relationship between
        $E(2190\Angstrom-2500\Angstrom)$ 
        and the central absorption at 2175$\Angstrom$,
        they took the bump ``equivalent width'' to be
  $\approx 160\Angstrom\magni^{-1}\times E(2190\Angstrom-2500\Angstrom)$.
        }
They found that the ``equivalent widths'' of the bump
and the central depths of the $\lambda$4430$\Angstrom$ DIB
were well correlated, with the correlation coefficient exceeding 
$\simali$0.9.
  
Schmidt (1978) compared $E(2200\Angstrom-3320\Angstrom)$ 
with the EWs of the $\lambda$5780$\Angstrom$ DIB
in the spectra of 50 early-type stars. 
He found that they were correlated, 
although with considerable intrinsic scatter.

Danks (1980) determined the ``depths'' of the 2175$\Angstrom$ 
bump from $D_{2175} \equiv A_{2175\Angstrom} 
- 0.5\times \left(A_{2500\Angstrom} + A_{1800\Angstrom}\right)$ 
of 30 stars, where $A_\lambda$ is the extinction 
at wavelength $\lambda$. 
He found a quite tight linear correlation between $D_{2175}$ 
and the EW of the $\lambda$4430$\Angstrom$ DIB.

Dorschner et al.\ (1977) estimated the 2175$\Angstrom$ bump
      strengths of sightlines toward 194 stars from integrating 
      the bump area 
      (with the continuum extinction subtracted).\footnote{% 
         They approximated the continuum extinction 
         underneath the 2175$\Angstrom$ bump by 
         a straight line between $\lambda^{-1} = 3\mum^{-1}$ 
         and $\lambda^{-1} = 6\mum^{-1}$. 
         } 
They found a fairly good correlation 
between the 2175$\Angstrom$ bump strength
and the central depth of the $\lambda$4430$\Angstrom$ DIB
(with a correlation coefficient of $\simali$0.74).\footnote{%
      Wu et al.\ (1981) argued that the bump strengths calculated 
      by Dorschner et al.\ (1977) may probably have included
      continuous absorption not typically associated with the bump.
      }
Similar correlations were found for
the 2175$\Angstrom$ bump strength and 
the EW of the $\lambda$5780$\Angstrom$ DIB
(with a correlation coefficient of $\simali$0.81)
and that of the $\lambda$5797$\Angstrom$ DIB
(with a correlation coefficient of $\simali$0.74).

Wu et al.\ (1981) estimated the 2175$\Angstrom$ bump
      strengths from $S\equiv E(2200\Angstrom-V) 
      - 0.5\times\left[E(1800\Angstrom-V) - E(2500\Angstrom-V)\right]$
      for sightlines toward 110 hot stars.
      They found that the central depth of 
      the $\lambda$4430$\Angstrom$ DIB is well correlated 
      with the 2175$\Angstrom$ extinction bump
      (with a correlation coefficient of $\simali$0.83).
      Much weaker correlations were found for 
      the 2175$\Angstrom$ bump strength and 
      the EW of the $\lambda$5780$\Angstrom$ DIB
      (with a correlation coefficient of $\simali$0.66)
      and that of the $\lambda$6284$\Angstrom$ DIB
      (with a correlation coefficient of $\simali$0.63).
%
%      They argued that the bump strengths calculated by
%      Dorschner et al.\ (1977) may probably have included
%      continuous absorption not typically associated with the bump.

Witt et al.\ (1983) first pointed out that the possible
correlation between the 2175$\Angstrom$ bump 
and the $\lambda$4430$\Angstrom$ DIB suggested by
Nandy \& Thompson (1975) and Wu et al.\ (1981) 
may merely reflect the fact that both the bump strength
and the DIBs correlate with $E(B-V)$.
In order to eliminate the common correlation with $E(B-V)$,
Witt et al.\ (1983) compared the central depths of 
the $\lambda$4430$\Angstrom$ DIB normalized by $E(B-V)$ 
with the normalized bump strengths
$E({\rm bump})/E(B-V)$ of 20 stars.\footnote{% 
  Witt et al.\ (1983) defined $E({\rm bump})$ 
  as the excess extinction at $\lambda^{-1} = 4.62\mum^{-1}$ 
  above a linear interpolation of the extinction curve 
  between $\lambda = 3.50\mum^{-1}$ and $5.75\mum^{-1}$.
  }
They only found a marginally significant
correlation between the normalized strengths of the UV bump
and the $\lambda$4430$\Angstrom$ DIB,
with a correlation coefficient of $\simali$0.52.
This was confirmed by Seab \& Snow (1984) who explored
a larger sample of 50 stars.
%%
%Ever since the work of Witt et al.\ (1983) and Seab \& Snow (1984),
%astronomers always compare the normalized strengths 
%of the extinction bump and DIBs.

Benvenuti \& Porceddu (1989) compared the EWs of six DIBs 
($\lambda$5780, 5797, 6196, 6203, 6270 and 6284$\Angstrom$)
with the 2175$\Angstrom$ bump strengths of a sample of 
26 galactic stars, with both quantities normalized by $E(B-V)$.
They took the bump height $h_{2175}$ (see Fig.\,1 and \S3)
for the bump strength and found no correlation 
between the bump and DIBs
(even for the closest correlation
[for the 2175$\Angstrom$ bump and the $\lambda$6284$\Angstrom$ DIB]
the correlation coefficient is only $\simali$0.39).
%which was measured using the prescription of 
%Fitzpatrick \& Massa (1990; hereafter FM90),
%i.e. as the extinction at $\lambda^{-1}=4.62\mum^{-1}$
%with respect to a linear ``background extinction''
%intersecting the curve at $\lambda^{-1}=3.5\mum^{-1}$
%and $\lambda^{-1}=5.75\mum^{-1}$ (see Fig.\,1 and \S3).
%They found no correlation between the bump and DIBs.
%
%After canceling the intrinsic correlation 
%due to the generally constant gas-to-dust ratio 
%of the interstellar medium, no further correlation 
%between the UV bump and the observed DIBs was found, 
%suggesting that dust and the carrier(s) of the DIBs, 
%although coexisting in the interstellar medium, 
%have an independent history. 
%
D\'esert et al.\ (1995) performed a similar analysis
for eight DIBs ($\lambda$5707, 5780, 5797, 5850, 6177, 
6196, 6269, 6284$\Angstrom$) of 28 stars.
In contrary to Benvenuti \& Porceddu (1989),
they found that the 2175$\Angstrom$ bump height
correlates with the DIB EWs,
with the correlation coefficients greater
than 0.8 for some DIBs.
More recently, Megier et al.\ (2005) compared the normalized
EWs of 11 DIBs with the 2175$\Angstrom$ bump strengths of 49 stars,
also using the bump height as a measure of the bump strength. 
They found that most of the DIBs correlate 
positively with the extinction bump.

\section{How to Characterize the 2175$\Angstrom$ 
         Extinction Feature Strength?}
In view of the various standards in determining 
the strength of the 2175$\Angstrom$ bump (see \S2),
an important question one need to address is:
what is the most reasonable measure of the strength 
of the 2175$\Angstrom$ bump?
We argue that the area of the continuum-subtracted 
bump integrated over $\lambda^{-1}$ is an appropriate 
measure (see below and Fig.\,1).

Fitzpatrick \& Massa (1990; hereafter FM90)
found that the 2175$\Angstrom$ extinction bump,
with the underneath linear background subtracted, 
can be closely fitted by a Drude profile.
The Drude profile, 
expected for classical damped harmonic oscillators,
is characterized by 
$\lambda_{\rm o}$ -- the peak wavelength, 
$\gamma$ -- the band width, 
and $c_3$ -- a parameter relates to the strength of the bump.
In the Galactic ISM, the strength and width of
the 2175$\Angstrom$ extinction bump vary with
environment while its peak position 
$\lambda_{\rm o} \approx 2175\Angstrom$
is quite invariant (see Whittet 2003).

With the extinction curve expressed as  $E(\lambda-V)/E(B-V)$,
where $E(\lambda-V)$ is the reddening between wavelength $\lambda$ 
and the visual band $V$, 
as illustrated in Fig.\,1, 
the area in the 2175$\Angstrom$ bump
above the linear background 
(i.e. the shaded area in Fig.\,1) 
for an extinction curve
{\it normalized by} $E(B-V)$ is
$W_{2175}/E(B-V) = \pi\,c_{3}/2\gamma$
(see Fitzpatrick \& Massa 1986, 1990).
The bump height, defined as the maximum height above 
the linear background, is 
$h_{2175}/E(B-V) = c_3/\gamma^2$.
We argue that $W_{2175}$ 
(which is analogous to the concept of equivalent width) 
is the most reasonable measure 
of the strength of the 2175$\Angstrom$ extinction bump, 
particularly in view of the fact that we use the equivalent width 
of a DIB to characterize its absorption strength, 
which measures the area of the DIB absorption profile 
above the continuum 
(see the shaded area in the inserted panel in Fig.\,1). 
%%%%
We note that both quantities linearly determine 
the column densities of their respective carriers:
as the DIBs are always optically thin, 
the equivalent width $W_{\rm DIB}$ of a DIB 
at wavelength $\lambda$ allows us to determine 
the column density of its carrier through
$N_{\rm DIB} \approx 1.13\times 10^{20}\cm^{-2} 
\left(W_{\rm DIB}/\Angstrom\right)
\left(\lambda/\Angstrom\right)^{-2}\,\fdib^{-1}$,
where $\fdib$ is the oscillator strength of the transition 
associated with the DIB (see Herbig 1993).
For the 2175$\Angstrom$ extinction bump, 
the column density of its carrier $N_{\rm bump}$ 
is also linearly related to $W_{2175}$:
$N_{\rm bump} = \left(m_e c/\pi e^2\right)\,\fbump^{-1}\,W_{2175}$,
where $\fbump$ is the oscillator strength (per absorber) 
associated with the 2175$\Angstrom$ bump.\footnote{%
  The column density $N_{\rm bump}$ of the carrier
  of the 2175$\Angstrom$ extinction bump 
  is linearly proportional to the total area of the bump 
  above the continuum extinction:
  $N_{\rm bump}/N_{\rm H} = 1.086 \left(m_e c/\pi e^2\right)\,\fbump^{-1}
  \int \left(\Delta \tau_\nu/N_{\rm H}\right) d\nu$,
  where $N_{\rm H}$ is the hydrogen column density, 
  $m_e$ is the electron mass, $c$ is the speed of light,
  $e$ is the electron charge, and
  $\Delta \tau_\nu/N_{\rm H}$ is the optical depth 
  (per H column) attributed to the bump (see Draine 1989).
  Since $\int \left(\Delta \tau_\nu/N_{\rm H}\right) d\nu = 
  \left(1/1.086\right)\,\left(\pi c_3/2\gamma\right)\,E(B-V)/N_{\rm H}$
  [where $E(B-V)/N_{\rm H}\approx 
   1.7\times 10^{-22}\magni\cm^2\,{\rm H^{-1}}$ 
   is the extinction-to-hydrogen ratio in the diffuse ISM],
   we obtain
   $N_{\rm bump} = \left(m_e c/\pi e^2\right)\,\fbump^{-1}\,W_{2175}$.
   % \approx 37.7\cm^{-2}\,f^{-1} \left(W_{2175}/\mum^{-1}\right)$.
   }

The bump strength relates to the bump height through 
$W_{2175} = \left(\pi\gamma/2\right)h_{2175}$.
Since the bump width $\gamma$ varies from one sightline to another, 
the bump height itself is not sufficient to describe the bump strength.

With $c_3$ and $\gamma$ determined from fitting the observed 
extinction curve with the FM90 decomposition,\footnote{%
   Fizpatrick \& Massa (1990) found that the interstellar 
   extinction curves at $3.3\mum^{-1} < \lambda^{-1} < 10\mum^{-1}$ 
   can be decomposed into three components 
   consisting of six parameters --- 
   (i) a linear ``background'' term;
   (ii) a ``Drude profile'' term 
        for the 2175$\Angstrom$ extinction feature; and
   (iii) a far ultraviolet (UV) non-linear rise at 
         $\lambda^{-1}>5.9\mum^{-1}$.
    }
one can obtain the bump strength (normalized by $E(B-V)$)
from $c_3$ and $\gamma$.
If $c_3$ is unknown, one can still obtain the bump strength simply 
from the bump height $h_{2175}$ and width $\gamma$.

\section{Correlation Between DIBs and 
         the 2175$\Angstrom$ Extinction Bump}
As a first step in examining whether the 2175$\Angstrom$ 
extinction bump correlates with the DIBs, we compile 
a large number of sightlines for which both the interstellar 
extinction curves and DIBs have been measured 
(see Tables~1,2 and the Appendix).
In this work we focus on nine strong DIBs 
in the visible wavelength range for which
rich sets of high quality spectroscopy data
are available: 
$\lambda$5707$\Angstrom$, 
$\lambda$5780$\Angstrom$, 
$\lambda$5797$\Angstrom$, 
$\lambda$6195\AA/6196$\Angstrom$, 
$\lambda$6203$\Angstrom$,
$\lambda$6269\AA/6270$\Angstrom$, 
$\lambda$6284$\Angstrom$, 
$\lambda$6376\AA/6379$\Angstrom$, and
$\lambda$6614$\Angstrom$. 
We group the two DIBs at $\lambda$6376$\Angstrom$ 
and $\lambda$6379$\Angstrom$ into one band 
(which is labeled ``$\lambda$6376\AA/6379$\Angstrom$'')
as they may blend.
%(see discussions later in this section).
%This is also done for the DIBs at
%$\lambda$6376$\Angstrom$ and $\lambda$6379$\Angstrom$.
The DIBs at $\lambda$6195$\Angstrom$ and $\lambda$6196$\Angstrom$
are actually the same DIB, which we refer as 
the ``$\lambda$6195\AA/6196$\Angstrom$''. 
Similarly, the ``$\lambda$6269\AA/6270$\Angstrom$'' DIB
refers to either the $\lambda$6269$\Angstrom$ DIB 
or the $\lambda$6270$\Angstrom$ DIB
(which are also the same DIB). 
Unfortunately, the equivalent width information 
is not always available for the strongest DIB which peaks 
at $\lambda$4430$\Angstrom$, in the literature most often 
only the absorption depth has been reported for this DIB. 

The DIB EWs and the extinction parameters 
($c_3$, $\gamma$) are taken from 
Dorschner et al.\ (1977), 
Fitzpatrick \& Massa (1986, 1990),
Benvenuti \& Porceddu (1989), 
Herbig (1993),
Jenniskens \& Greenberg (1993),
D\'esert et al.\ (1995), 
Sonnentrucker et al.\ (1997), 
Thorburn et al.\ (2003), Galazutdinov et al.\ (2004), 
Lewis et al.\ (2005), Megier et al.\ (2005), 
Cox et al.\ (2007b), Hobbs et al.\ (2008, 2009),
McCall et al.\ (2010), and Friedman et al.\ (2011).\footnote{%
    Thorburn et al.\ (2003) reported the EWs of nine DIBs at
    $\lambda$5780$\Angstrom$, $\lambda$5797$\Angstrom$, 
    $\lambda$6196$\Angstrom$, $\lambda$6203$\Angstrom$, 
    $\lambda$6270$\Angstrom$, $\lambda$6284$\Angstrom$, 
    $\lambda$6376$\Angstrom$, $\lambda$6379$\Angstrom$, 
    and $\lambda$6614$\Angstrom$ 
    (in their paper they labeled it $\lambda$6613$\Angstrom$). 
    More recently, those authors updated the EWs of five of these DIBs:
    $\lambda$5780$\Angstrom$, $\lambda$5797$\Angstrom$, 
    $\lambda$6196$\Angstrom$, $\lambda$6284$\Angstrom$, 
    and $\lambda$6614$\Angstrom$ (see Friedman et al.\ 2011).
    We therefore adopt the EWs of the four DIBs 
    ($\lambda$6203$\Angstrom$, $\lambda$6270$\Angstrom$,
     $\lambda$6376$\Angstrom$, and $\lambda$6379$\Angstrom$) 
     of Thorburn et al.\ (2003),
    and the EWs of the five DIBs   
    ($\lambda$5780$\Angstrom$, $\lambda$5797$\Angstrom$, 
    $\lambda$6196$\Angstrom$, $\lambda$6284$\Angstrom$, 
    and $\lambda$6614$\Angstrom$)
    of Friedman et al.\ (2011).
%
%   {\bf  The authors rely on literature values of the EWs of the DIBs. 
%   They should note for the reader that the issue of what limits 
%   of integration are used for calculating the DIB EWs has caused 
%   many problems, especially as the number of DIBs increases. 
%   Some authors include certain blends as a single DIB, 
%   others do not. 5797 is a good example.
%   Galazutdinov et al.\ (2004, MNRAS, 355, 169) 
%   do not include the blue wing, while in a recent paper 
%   Friedman et al.\ (2011, ApJ, 727, 33) do include it. 
%   This difficulty should be mentioned.
   }
In Table~1 we tabulate for each object the extinction parameters 
$E(B-V)$, $R_V$, $c_3$, $\gamma$, and $W_{2175}/E(B-V)$,
as well as the EWs of the nine DIBs. 
In the Appendix we elaborate how we obtain
the EWs of these DIBs. 

%  We see in these tables that some sources are listed
%  more than once (e.g., HD\,2095, HD\,34078, HD\,37061). 
%  This is because these sightlines have been 
%  observed more than once (in different epochs
%  by different authors).
%  In the correlation analysis, we take the mean values.

In Figure 2 we show the correlation between 
$\WDIBp \equiv W_{\rm DIB}/E(B-V)$,
the normalized DIB EWs, and 
$\Wbumpp \equiv W_{2175}/E(B-V) = \pi c_3/2\gamma$, 
the normalized bump strengths.\footnote{%
  We normalize both $W_{2175}$ and $W_{\rm DIB}$ 
  by $E(B-V)$ to cancel out the linear correlations of
  $E(B-V)$ with $W_{\rm DIB}$ and $W_{2175}$
  which reflect the amount of interstellar materials. 
  $W_{\rm DIB}$ is the EW of a DIB at a specific wavelength. 
  }
It is clearly seen in this figure 
that the correlation is very poor.
To be more quantitative, we calculate the correlation coefficients 
$r$ of the bump strengths and the DIB EWs,
both are normalized to $E(B-V)$:
$r = \left\{\langle\Wbumpp\WDIBp\rangle
            -\langle\Wbumpp\rangle\langle\WDIBp\rangle
      \right\}/\left\{\sigma_{\Wbumpp} \sigma_{\WDIBp}\right\}$,
where $\sigma_{\Wbumpp}$ and $\sigma_{\WDIBp}$
are the variances of $\Wbumpp$  and $\WDIBp$, respectively. 
Table~2 presents the calculated correlation coefficients.
We see that $r<0.60$ for all DIBs, indicating that 
the 2175$\Angstrom$ bump is poorly if not at all
correlated with any of the nine DIBs
(even for the most ``correlated'' one, $\lambda$6614$\Angstrom$,
the correlation coefficient is only $r\approx 0.58$).\footnote{%
  Although somewhat arbitrary, we feel a close correlation
  should have $r>0.9$, a good correlation should have $r>0.8$,
  and a fair correlation should have $r>0.7$.
  This seems to be in line with previous studies (see \S2).
  In studying the correlation between the 2175$\Angstrom$ 
  extinction bump and the far-UV extinction, 
  Greenberg \& Chlewicki (1983) described the correlation 
  with $r=0.731$ as being ``poor''.
  }

\section{Discussion}\label{sec:discussion}
A novel aspect of the present study is that, we measure
the bump strength as the area above the linear background
(see Fig.\,1) instead of the height. Since the DIB EW is
essentially a measure of the area of the DIB line profile,
it seems more appropriate to us to compare the DIB EW with 
the bump area rather than the bump height. The bump area 
differs from the bump height by a factor of $\pi\gamma/2$.
Note that although the central wavelength of 
the 2175$\Angstrom$ bump is stable,
its width $\gamma$ varies substantially (see Whittet 2003):
while on average $\gamma\approx 1.0\mum^{-1}$,
$\gamma$ ranges from $\gamma<0.8\mum^{-1}$ 
(e.g. $\gamma\approx0.79\mum^{-1}$ for HD\,93028)
to $\gamma>1.2\mum^{-1}$ 
(e.g. $\gamma\approx1.25\mum^{-1}$ for $\xi$ Oph
and $\gamma\approx1.62\mum^{-1}$ for HD\,29647).

Does the lack of correlation between the strengths 
of the 2175$\Angstrom$ bump and DIBs (see \S3) 
imply that their carriers are not related 
as often suggested in the literature
(e.g. see Wu et al.\ 1981, Witt et al.\ 1983, 
Seab \& Snow 1984, Benvenuti \& Porceddu 1989)?
or in other words, does this conclusively imply
that PAHs are unlikely responsible for 
both the 2175$\Angstrom$ bump and DIBs?

Cox et al.\ (2007a) studied the 2175$\Angstrom$ extinction bump
and DIBs of the SMC. They reported the detection of DIBs toward 
one line of sight in the relatively quiescent SMC wing region 
where the 2175$\Angstrom$ extinction feature is also present. 
In contrast, neither DIBs nor the 2175$\Angstrom$ bump were 
detected toward five lines of sight in the SMC bar which pass 
through active star formation regions. They argued that the carriers 
of the 2175$\Angstrom$ extinction bump and DIBs may be subject to 
the same physical and chemical processes: 
both carriers in the SMC bar could be destroyed by shocks 
or the hard ambient UV radiation; 
or alternatively, the harsh environment may prohibit 
the formation of both carriers. 
We note that in the Galaxy, some lines of sight also exhibit 
both weak DIBs and a weak extinction bump
(e.g. HD\,23532, Witt et al.\ 1981; HD\,29647, Adamson et al.\ 1991; 
      HD\,62542, Snow et al.\ 2002, \'Ad\'amkovics et al.\ 2005)
on a per unit reddening basis, when compared with typical
Galactic sightlines. 

One may argue that it may not be appropriate to compare 
the EW of a single DIB with the 2175$\Angstrom$ bump 
since an individual DIB is more characteristic of 
a specific PAH molecule, while the 2175$\Angstrom$ bump 
is unlikely from a single PAH molecule, 
but rather produced by a cosmic mixture of 
many individual molecules, radicals, and ions. 
In view of this, we compare the sum of the EWs of 
all nine DIBs $\sum_{j=1}^{N=9} W_{{\rm DIB}_j}$ 
with the 2175$\Angstrom$ bump strengths $W_{2175}$ 
of nine sightlines (for which the data for both 
the nine DIBs and the bump are available),
where $W_{{\rm DIB}_j}$ is the EW of the $j$-th DIB.
As shown in Fig.\,3, the correlation between
$\sum_{j=1}^{N=9} W_{{\rm DIB}_j}$ and $W_{2175}$ 
is at most very weak. We do not want to over-interpret
this as  
%there does not appear to exist 
%any strong correlation between 
%$\sum_{j=1}^{N=9} W_{{\rm DIB}_j}$ and $W_{2175}$ either. 
the scatter is rather large and there is only a small 
number of data points. For a more thorough exploration,
we need to include more sightlines (i.e. more data points)
and an as complete as possible set of DIBs. The latter will
allow us to approach the total absorption of DIBs.

PAHs are thought to be ubiquitous and abundant in the ISM, 
as revealed by the ubiquitous ``UIR'' emission bands at 
3.3, 6.2, 7.7, 8.6 and 11.3$\mum$ 
(L\'eger \& Puget 1984, Allamandola et al.\ 1985)
which account for over 10\% of the total IR luminosity 
of the Milky Way and external star-forming galaxies 
(Draine \& Li 2007, Smith et al.\ 2007). 

PAHs do in general have strong absorption in 
the $\simali$2000$\Angstrom$ region
as a result of the strong $\pi$--$\pi^{\ast}$ 
electronic transition. It is therefore natural 
to attribute at least part of the 2175$\Angstrom$ 
extinction feature to PAHs (see Draine 2009). 
Indeed, a correlation between the 2175$\Angstrom$ hump 
and the IRAS 12$\mum$ emission (dominated by PAHs) was 
found by Boulanger et al.\ (1994) 
in the Chamaeleon cloud, suggesting a common carrier.

While it is true that a single PAH molecule exhibits sharp UV 
absorption features which are not seen in interstellar 
space, a cosmic mixture of various individual PAH molecules, 
radicals, and ions would smooth out the fine structures and 
produce a broad absorption bump around
$\simali$2175$\Angstrom$. This has been demonstrated 
both experimentally and theoretically.
L\'eger et al.\ (1989) measured the absorption spectra
of mixtures of over 300 neutral PAH species 
with $\simali$12--28 C atoms
obtained from a coal pitch 
extract evaporated at $\simali$380--480\,K. 
Joblin et al.\ (1992) measured the absorption 
spectra of the neutral PAH mixtures containing 
$\simali$14--44 C atoms obtained from a coal pitch extract 
evaporated at $\simali$570--630\,K.
All these spectra have a strong UV feature around 2175$\Angstrom$
(but relatively broader than the interstellar bump).
Cecchi-Pestellini et al.\ (2008) recently showed 
that a weighted sum of 50 PAHs in the size range of
$\simali$10--66 C atoms in the charge states of 0 (neutral),
$\pm1$ (cation and anion), and $+$2 (dication)
is able to reproduce the 2175$\Angstrom$ extinction bump
observed in five sightlines with $R_V$ ranging from
2.33 to 5.05 (also see Malloci et al.\ 2008).

%Indeed, both experimentally-measured 
%and quantum-chemically-calculated spectra 
%of complex mixtures of many individual PAH molecules 
%exhibit an absorption bump resembling 
%the 2175$\Angstrom$ interstellar extinction feature32-35. 

So, why do not DIBs correlate with the extinction bump?
A possible explanation might be that DIBs are produced 
by the smallest gas-phase free-flying PAH molecules and ions, 
while the 2175$\Angstrom$ bump is mainly from large PAHs
or PAH clusters in condensed phase. 
Therefore, one should not expect the DIBs to be tightly 
correlated to the 2175$\Angstrom$ bump.
Indeed, models that reproduce 
the $\simali$3--20$\mum$ ``UIR'' emission bands 
have most of the PAH mass in PAHs 
with $N_{\rm C}\gtsim\,10^{2}$ C atoms 
(Li \& Draine 2001, Draine \& Li 2007)\footnote{%
   This scenario naturally explains the lack of correlation
   between the strengths of the $\lambda$5780$\Angstrom$
   and $\lambda$6283$\Angstrom$ DIBs with the {\it Spitzer}
   8$\mum$ IRAC fluxes (which are thought to arise from PAHs)
   observed in the ISM of M31 (Cordiner et al.\ 2011): 
   small PAHs (say, with $N_{\rm C}$\,$<$\,30--50 C atoms) 
   which are responsible for the DIBs do not emit much 
   at the 7.7$\mum$ C--C stretching mode
   (instead, they probably mainly emit at the 3.3$\mum$ 
    C--H stretching mode), only large PAHs 
    with $N_{\rm C}\gtsim\,10^{2}$ C atoms dominate
    the 7.7$\mum$ emission (see Fig.\,7 of Draine \& Li 2007).
    }
while laboratory studies are generally
limited to smaller PAHs
(e.g. $12 < N_{\rm C} <28$ in L\'eger et al.\ 1989,
and $14< N_{\rm C} <44$ in Joblin et al.\ 1992). %\footnote{%
  This may also be the reason why
  the PAH samples in the laboratory 
  (L\'eger et al.\ 1989, Joblin et al. 1992)
  do not precisely reproduce the observed profile 
  of the 2175$\Angstrom$ feature: most notably
  is the feature at $\simali$3000$\Angstrom$ which
  is seen in the laboratory spectra but not observed 
  in the ISM. Joblin et al.\ (1992) noted that this 
  feature shifts in the laboratory spectra
  towards longer wavelengths as molecules 
  become larger. Joblin et al.\ (1992) argued that 
  one would expect that this feature 
  merges into a continuum when larger species
  are present in the laboratory mixture. 
  More recently, Steglich et al.\ (2010) showed that larger PAHs 
  indeed provide better fits to the observed 2175$\Angstrom$ feature.
  %}
%
%However, as experimentally demonstrated by 
%L\'eger et al.\ (1989; $12 < N_{\rm C} <28$) and 
%Joblin et al.\ (1992; $14< N_{\rm C} <44$), 
%mixtures of small free gas-phase neutral PAHs  
%do show a broad bump around 2175$\Angstrom$ 
%in their absorption spectra.
%
Alternatively, DIBs may mainly arise from
small PAH ions while the 2175$\Angstrom$ hump 
is weakened in cations (e.g. Lee \& Wdowiak 1993). 
Moreover, PAHs may not be the sole contributor 
to the 2175$\Angstrom$ bump. Other dust species such as 
small graphitic dust and carbon buckyonions may also 
(at least partly) contribute to the bump (see Li et al.\ 2008). 
Therefore, although PAHs may indeed be responsible 
for both the extinction bump and DIBs, their strengths 
do not have to correlate 
(i.e. they do not necessarily have 
a one-to-one correspondence). 

Finally, if DIBs are indeed caused by the electronic
transitions of PAHs, they hold great promise for 
identifying specific PAH molecules.
So far, not a single individual PAH molecule has 
been identified in the ISM,\footnote{%
  The naphthalene cation (C$_{10}$H$_8^{+}$)
  was tentatively suggested to be responsible
  for several DIBs (Salama \& Allamandola 1992, Snow 1992,
  Kre{\l}owski et al.\ 2001, Iglesias-Groth et al.\ 2008).
  Iglesias-Groth et al.\ (2010) also reported a tentative 
  detection of anthracene cation (C$_{14}$H$_{10}^{+}$).
  However, these claimed detections were not confirmed 
  by Galazutdinov et al.\ (2011).
  }
despite that many unidentified interstellar phenomena 
(e.g. the 2175$\Angstrom$ bump, DIBs,
the so-called blue luminescence [Vijh et al.\ 2005a,b]
and extended red emission [ERE; Witt \& Vijh 2004], 
the 3.3--11.3$\mum$ emission bands, and
the ``anomalous'' Galactic foreground microwave emission
[Draine 2003])
are arguably attributed to PAHs.
Neither the 2175$\Angstrom$ extinction bump 
nor the 3.3--11.3$\mum$ ``UIR'' emission features 
could allow us to fingerprint individual PAH molecules 
as they are caused by a mixture of PAH molecules of 
a range of sizes. Moreover, the mid-IR bands are mostly 
representative of functional groups, not the molecular frame 
of a PAH molecule.
In contrast, the DIBs and the far-IR vibrational bands are 
sensitive to the skeletal characteristics of a PAH molecule, 
hence they contain specific fingerprint information on its 
identity (Salama et al.\ 1996, 1999; Joblin et al.\ 2009, 
Zhang et al.\ 2010).

%If I had to guess I would suspect the diffuse bands 
%have a gas-phase origin and 2175 a condensed phase origin. 
%But the DIB carriers could well be linked with the 2175 
%carrier(s), which, perhaps when eroded, 
%generate the diffuse band carriers.
%
%... the bump is caused by a solid form of aromatic carbon 
%(i.e, with a band gap energy) and that gas-phase PAHs may 
% contribute at the most account for 10\% of the extinction 
%(if at all). 
%Joblin et al. invoke the high energy pi-pi transitions of 
%neutral PAHs in the UV for the bump while the DIBs invoke 
%the pi-pi transitions of PAH ions in the visible and/or 
%the low energy pi-pi transitions of neutral PAHs 
%(in the visible). 
%
%
%Indeed, models that employ PAHs to reproduce 
%the $3-30\mum$ emission from interstellar dust 
%in galaxies with metallicities similar to 
%the Milky Way employ PAH abundances that are 
%sufficient to account for the strength of the
%2175 $\Angstrom$ feature. 

\section{Summary}\label{sec:summary}
We have explored the relationship between the 2175$\Angstrom$
interstellar extinction bump with nine DIBs, using the area
above the linear extinction background as a measure of the
bump strength. We have compiled the extinction bump and DIB 
equivalent width data of a large sample of 84 stars. 
It is found that the DIB EWs are not correlated with
the 2175$\Angstrom$ extinction bump.
A possible explanation might be that DIBs are produced 
by small free gas-phase PAH molecules and ions, 
while the 2175$\Angstrom$ bump is mainly from large PAHs
or PAH clusters in condensed phase. 

%We stress that this does not readily rule out 
%the hypothesis that PAHs are responsible for
%both the extinction bump and DIBs (although it either does 
%not provide favourable support) since the former
%may result from a cosmic mixture of various PAH
%species, radicals and ions, while the latter is
%more likely from a specific individual PAH molecule.

\acknowledgments{We thank L.J. Allamandola, S.D. Friedman,
J. Gao, J.Y. Hu, S. Kwok, S.L. Liang, B.J. McCall,
F. Salama, P.J. Sarre, and A.N. Witt 
for very helpful discussions and/or comments.
We also thank the anonymous referee for his/her 
very helpful comments and suggestions which substantially
improved the quality of this work.
FYX is supported in part by the Scientific Research Fund 
of Hunan Provincial Education Department (No.\,08A072), 
NSFC grant No.\,10878012, the Xiangtan University Natural 
Science Foundation (No.\,10KZ08044) and the Key Laboratory of 
Particle Astrophysics, Institute of High Energy Physics, CAS.
AL is supported in part by Spitzer Theory
Programs and NSF grant AST 07-07866.}

%%%%%%%%%%%%%%%%%%%%%%%%%%% Appendix %%%%%%%%%%%%%%%%%%%%%%%%%%%%
\appendix
\section{The Equivalent Widths of Individual DIBs}
In the literature, different equivalent widths have 
often been reported for a single DIB of a given sightline
(see Tables 3--11).
In some cases the reported EWs differ
substantially from one another.\footnote{%
        In some cases this is (at least partly) caused
        by the different wavelength limits of integration
        used to derive the EW of a DIB.
        This becomes even more true 
        as the number of DIBs increases. 
        Some authors include certain blends as a single DIB, 
        others do not
        [e.g., in calculating the EW of 
        the $\lambda$5797$\Angstrom$ DIB,
        Galazutdinov et al.\ (2004) did not include 
        the blue wing, while Friedman et al.\ (2011)
        did include it]. 
   }
It is therefore not trivial to select 
the ``true'' EW of a DIB.
We employ the following criteria to adopt
the EW of a DIB of a given 
line of sight from the literature: 
(1) if the sightline has been observed only once,
    we will adopt the DIB EW reported from that observation;
(2) if the sightline has been observed twice,
    we will take the mean value of the reported EWs 
    if they are close (denoted by ``$\surd$'' in Tables 3--11);
    otherwise we will take the one
    which falls in the $W_{\rm DIB}$\,--\,$E(B-V)$
    linear relation\footnote{% 
      For a given DIB, we obtain the $W_{\rm DIB}$\,--\,$E(B-V)$ 
      linear relation by using the $W_{\rm DIB}$ and $E(B-V)$
      data of all 84 sightlines studied here. 
      A linear relation between $W_{\rm DIB}$ and $E(B-V)$       
      is expected as both quantities are proportional to
      the amount of interstellar materials along the line of sight.
     }
    and reject the one which deviates significantly 
    from that relation
    (denoted by ``$\times$'' in Tables 3--11);
(3) if the sightline has been observed more than twice
    and for a single DIB there are three or more sets 
    of DIB EWs reported in the literature, 
    we will take the mean value of those 
    reported EWs which are close to each other
    and reject those which deviate significantly 
    from others.
%Finally, we intend to favour the latest results. 

%  We see in these tables that some sources are listed
%  more than once (e.g., HD\,2095, HD\,34078, HD\,37061). 
%  This is because these sightlines have been 
%  observed more than once (in different epochs
%  by different authors).
%  In the correlation analysis, we take the mean values.

%%%%%%%%%%%%%%% References %%%%%%%%%%%%%%%%%%%%%%%%%%%%

%\bibliographystyle{apj}
%\bibliography{hd5612}

\clearpage

%\clearpage

%%%%%%%%%%%%%%%%%%%% Figure 1 %%%%%%%%%%%%%%%%%%%%%%%%%%%%%%
\begin{figure}
  \includegraphics[width=16cm,angle=0]{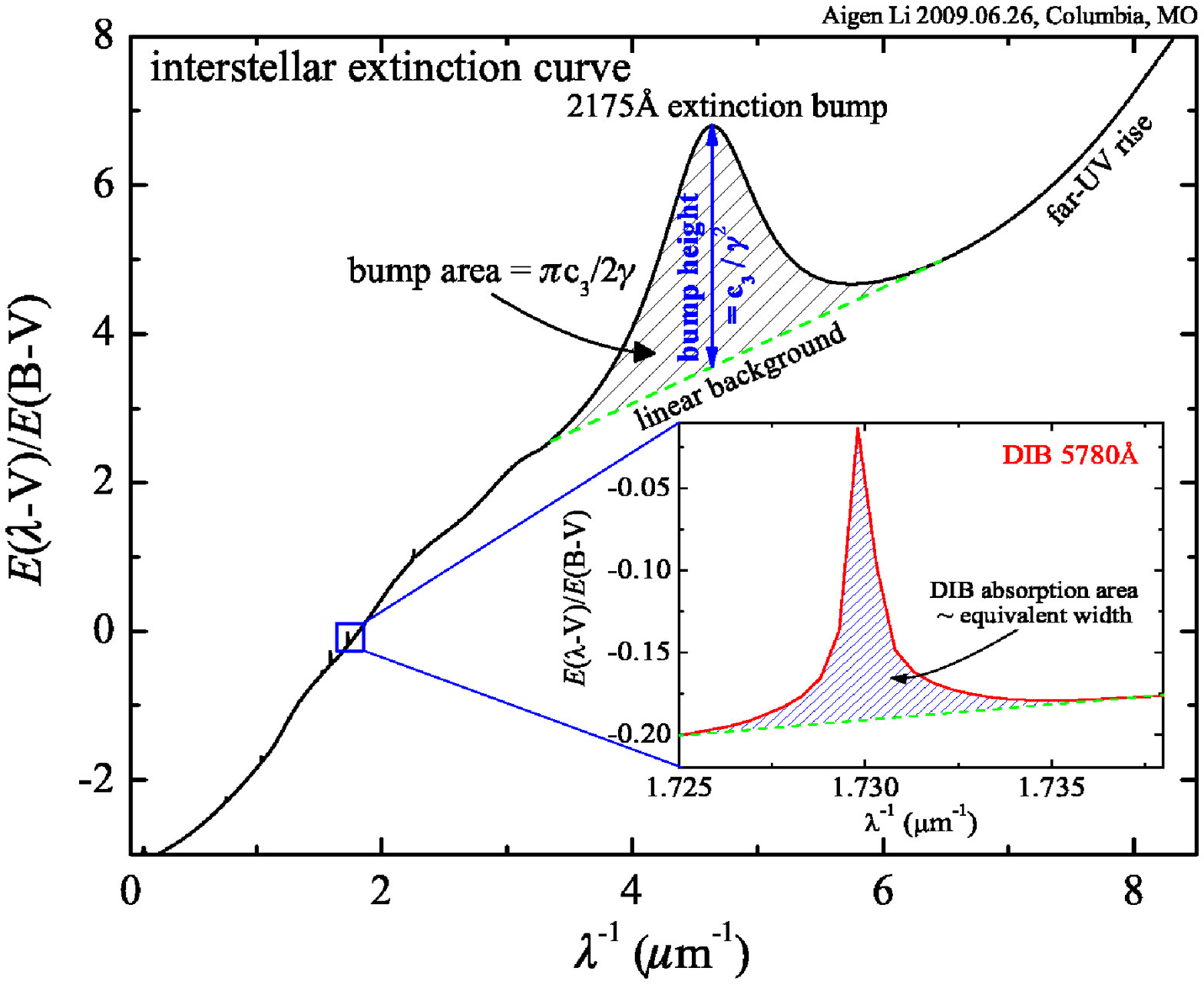} %\\[8mm]
  \caption{\footnotesize
           Interstellar extinction curve and DIBs.
           We use the black shaded area 
           ($W_{2175}/E(B-V)=\pi c_3/2\gamma$) above the
           ``linear background extinction'' in
            the FM90 decomposition scheme as a measure
            of the strength of the 2175$\Angstrom$
            extinction bump (normalized by $E(B-V)$), 
            while most previous work
            considered the bump height 
            ($h_{2175}/E(B-V)=c_3/\gamma^2$) 
            as the bump strength.  
            The little sticks in the visible part
            plot some DIBs. In the inserted panel which
            illustrates the $\lambda$5780$\Angstrom$ DIB,
            the blue shaded area represents its absorption
            strength.                        
            }
  \label{fig:1}
\end{figure}

%%%%%%%%%%%%%%%%%%% Figure 2 %%%%%%%%%%%%%%%%%%%%%%%%%%%%
\begin{figure}
\begin{center}
\includegraphics[width=16cm,angle=0]{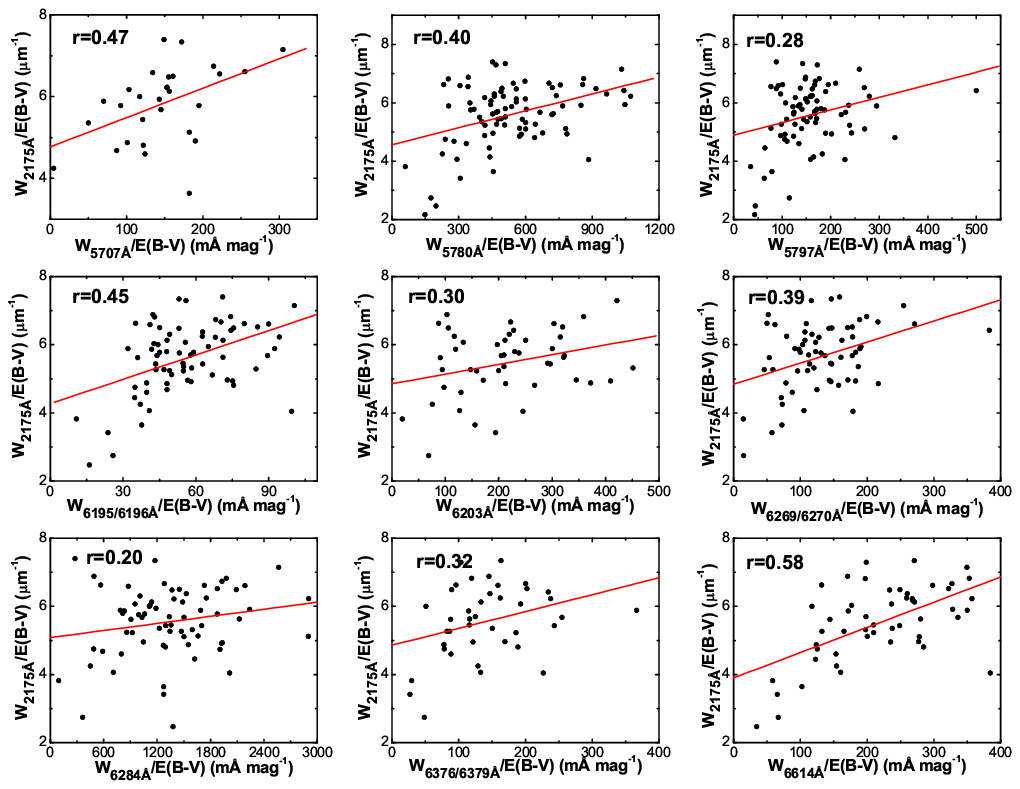}
\caption{\footnotesize\label{fig:fig2}
         Correlation diagrams of the strengths of 
         the 2175$\Angstrom$ extinction bump with 
         the equivalent widths of 
         the DIBs at $\lambda$5707$\Angstrom$, 
         $\lambda$5780$\Angstrom$,
         $\lambda$5797$\Angstrom$, 
         $\lambda$6195$\Angstrom$/6196$\Angstrom$, 
         $\lambda$6203$\Angstrom$,
         $\lambda$6269$\Angstrom$/6270$\Angstrom$, 
         $\lambda$6284$\Angstrom$,  
         $\lambda$6376$\Angstrom$/6379$\Angstrom$, 
         and $\lambda$6614$\Angstrom$.   
         The $\lambda$6195$\Angstrom$/6196$\Angstrom$ DIB 
         is a single DIB, although in the literature it is 
         listed either as $\lambda$6195$\Angstrom$ 
         or $\lambda$6196$\Angstrom$.
         This is also true for 
         the $\lambda$6269$\Angstrom$/6270$\Angstrom$ DIB. 
         The two DIBs at $\lambda$6376$\Angstrom$ 
         and $\lambda$6379$\Angstrom$ are grouped into one DIB 
         (``$\lambda$6376$\Angstrom$/6379$\Angstrom$'') 
         as they may blend.        
         }
\end{center}
\end{figure}

%%%%%%%%%%%%%%%%%%% Figure 3 %%%%%%%%%%%%%%%%%%%%%%%%%%%%
\begin{figure}
\begin{center}
\includegraphics[width=16cm,angle=0]{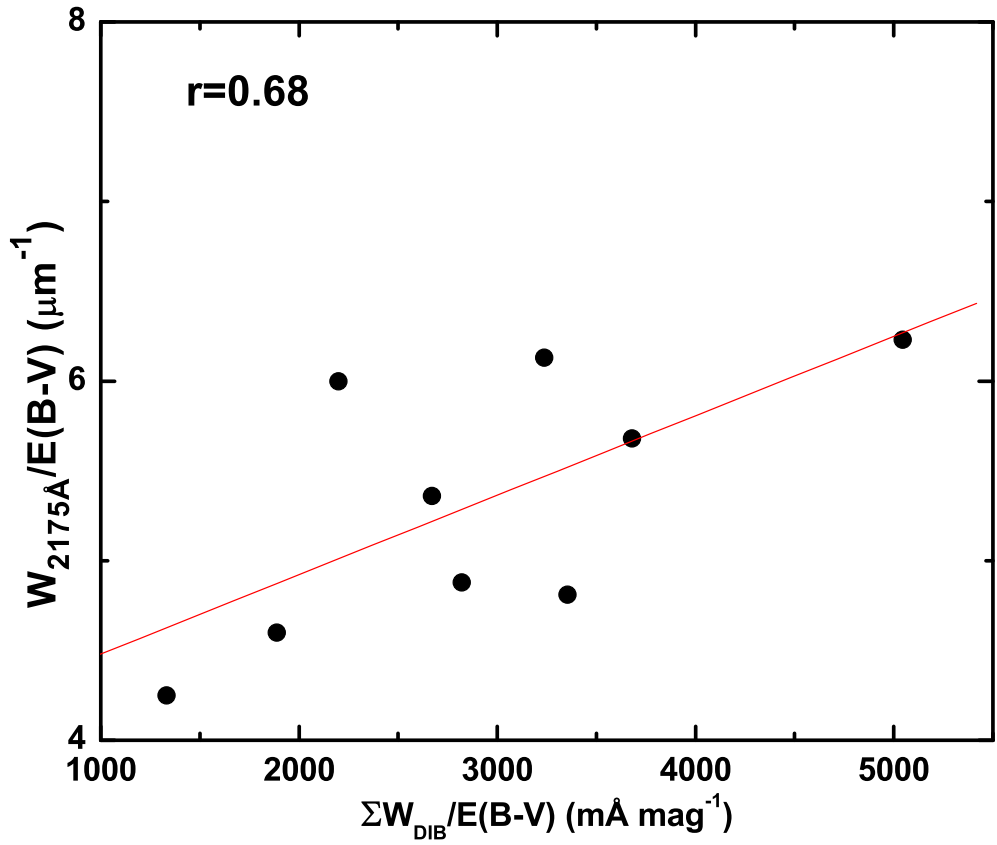}
\caption{\footnotesize\label{fig:fig3}
         Correlation diagram of the strengths of 
         the 2175$\Angstrom$ extinction bump with 
         the sum of the equivalent widths of all 
         nine DIBs
         ($\sum_{j=1}^{N=9} W_{{\rm DIB}_j}$)
         for nine interstellar sightlines.
         All quantities are normalized to $E(B-V)$.
         }
\end{center}
\end{figure}

%\clearpage

%%%%%%%%%%%%%%%%%%%%% Table 1 %%%%%%%%%%%%%%%%

\begin{landscape}
%
%\end{landscape}

\end{document}